\documentclass[amsmath,twocolumn,showpacs,showkeys]{revtex4}
\usepackage{amsmath,amsfonts,amssymb}

\def\a{\alpha}
\def\b{\beta}
\def\g{\gamma}
\def\vt{\vartheta}
\def\d{\delta}

\def\@{\partial_}

\def\negenspace{\kern-1.1em}

\def\sqr#1#2{{\vcenter{\hrule height.#2pt\hbox{\vrule width.#2pt
height#1pt \kern#1pt \vrule width.#2pt}\hrule height.#2pt}}}

\usepackage{graphicx, epic, eepic, color}

\definecolor{wichtig}{rgb}{1,0,0} 
\definecolor{folge}{rgb}{0,0,1} 
\definecolor{liste}{rgb}{0,0.7,0} 

\definecolor{dark-green}{rgb}{0,0.7,0}
\definecolor{dark-blue}{rgb}{0,0.2,0.5}
\definecolor{med-blue}{rgb}{0,0.7,1}
\definecolor{mblue}{rgb}{0,0.2,1}
\definecolor{cnc}{rgb}{0.8,0,0}
\definecolor{light-red}{rgb}{1,0.8,0.8}
\definecolor{dark-yellow}{rgb}{1,0.8,0}
\definecolor{light-blue}{rgb}{0.8,0.9,1}
\definecolor{verylight-blue}{rgb}{0.93,0.95,1}
\definecolor{light-yellow}{rgb}{1,0.9,0.8}
\definecolor{grey}{gray}{0.88}

\def\bfx{\mathbf{x}}

\def\bfy{\mathbf{y}}

\begin{document}
\title{Nonlocal Gravity Simulates Dark Matter}

\author{Friedrich W. Hehl}
\email{hehl@thp.uni-koeln.de} 
\affiliation{Institute for Theoretical Physics, University of Cologne,
50923 K\"oln, Germany}
\affiliation{Department of Physics and Astronomy,
University of Missouri-Columbia, Columbia, MO 65211, USA}
\author{Bahram Mashhoon}
\email{mashhoonb@missouri.edu}
\affiliation{Department of Physics and Astronomy,
University of Missouri-Columbia, Columbia, MO 65211, USA}

\begin{abstract}
  A nonlocal generalization of Einstein's theory of gravitation is
  constructed within the framework of the translational gauge theory
  of gravity. In the linear approximation, the nonlocal theory can be
  interpreted as linearized general relativity but in the presence of
  ``dark matter'' that can be simply expressed as an integral
  transform of matter. It is shown that this approach can accommodate
  the Tohline-Kuhn treatment of the astrophysical evidence for dark
  matter.

\end{abstract}



\date{03 February 2009, {\it file nlGrav12.tex}}

\maketitle

1.\ {\it Background.} In a recent paper \cite{BahramAnnalen}, the main
outlines of a nonlocal theory of special relativity have been sketched
and some of its experimental consequences have been described. The
extension of Lorentz invariance to accelerated observers in the
standard theory of relativity is based on a postulate of locality
that---on physical grounds---is expected to be just an approximation.
The recognition of the {\it limitations of the locality assumption}
has led to the development of nonlocal special relativity (see
\cite{BahramAnnalen,Bahram2007} and the references cited therein). It
is natural to consider the extension of this acceleration-induced
nonlocality to the gravitational field, as the principle of equivalence
establishes a connection between inertia and gravitation.  However, a
{\it direct} approach to a nonlocal generalization of general
relativity (GR) fails, since Einstein's principle of equivalence is
{\it strictly} local. Furthermore, the heuristic power of Einstein's
principle of equivalence cannot be employed to gain insight into the
physical connection between nonlocal gravity and nonlocal special
relativity. This circumstance provides the incentive to search for
other viable approaches to a nonlocal generalization of Einstein's
theory of gravitation.

2.\ {\it GR$_{||}$.} It turns out that general relativity can also be
obtained as a special case of the translational gauge theory of
gravity (see \cite{HehlHeld}; new developments and further references
are contained in recent papers \cite{Telecoll} as well as monographs
\cite{Milutin} and \cite{Ortin}). This teleparallel equivalent of
general relativity (GR$_{||}$) can be succinctly described within the
framework of a Weitzenb\"ock spacetime.

In Weitzenb\"ock geometry, the gravitational potential is represented
by a coframe 1-form $\vt^\a=e_i{}^\a dx^i$, where $e_i{}^\a(x)$ is the
tetrad field. Here $i, j,... = 0,\, 1,\, 2,\, 3$ denote holonomic
coordinate indices, while $\a,\b,... =\hat{0},\,\hat{1}, \,\hat{2},
\,\hat{3}$ denote anholonomic frame indices. We choose units such that
$c = 1$ throughout; moreover, the Minkowski metric tensor $\eta$ is
given by diag$(1, -1, -1, -1)$. The frame $e_\a=e^i{}_\a\partial_i$ is
dual to the coframe; therefore,
\begin{equation}\label{dual}
e_i{}^\a e^i{}_\b=\d_\b^\a\,,\quad e_i{}^\a e^j{}_\a=\d_i^j\,.
\end{equation}
The gravitational field strength is the object of anholonomity 2-form
$C^\a:=d\vt^\a$, where $C^\a=\frac 12 C_{ij}{}^\a dx^i\wedge dx^j$;
that is,
\begin{equation}\label{fieldstrength}
C_{ij}{}^\a=2\partial_{[i}e_{j]}{}^\a\,.
\end{equation}
We choose the frame field to be orthonormal throughout. The spacetime
interval is given by $ds^2=g_{ij}dx^i\!\otimes\! dx^j$, where
\begin{equation}\label{metric}
g_{ij}=\eta_{\a\b\,}e_i{}^\a  e_{j}{}^\b\,.
\end{equation}
A geodesic between two events is defined by the spacetime path that is
an extremum of the spacetime distance between them. Finally, in
Weitzenb\"ock geometry the frame field can be chosen to be globally
teleparallel. That is, the connection 1-form
$\Gamma_\a{}^\b=\Gamma_{i\a}{}^\b dx^i$ is such that the curvature
2-form
\begin{equation}\label{tele}
  R_\a{}^\b
  :=d\Gamma_\a{}^\b-\Gamma_\a{}^\g\wedge \Gamma_\g{}^\b 
=\frac 12 R_{ij\a}{}^\b dx^i\wedge dx^j
\end{equation}
is zero. Therefore, the frames in Weitzenb\"ock spacetime can be so
chosen that the connection vanishes everywhere,
\begin{equation}\label{gauge}
\Gamma_\a{}^\b\stackrel{*}{=}0\,.
\end{equation}
Here the star over the equality sign indicates that the equation is
valid only in the specially chosen global frames. Thus Weitzenb\"ock
spacetime has vanishing curvature but its torsion is in general
nonzero and in the special global frames reduces to the gravitational
field strength $C_{ij}{}^\a$.

3.\ {\it GR$_{||}$ field equations.} The gravitational field equations
of the translational gauge theory of gravity are given by
\cite{HehlHeld}
\begin{equation}\label{fieldeq}
  \partial_j{\cal H}^{ij}{}_\a-{\cal E}_\a{}^i\stackrel{*}{=}{\cal
    T}_\a{}^i\,,\quad \partial_{[i} C_{jk]}{}^\a\stackrel{*}{=}0\,,
\end{equation}
where ${\cal T}_\a{}^i=\sqrt{-g}\, T_\a{}^i$ and 
\begin{equation}\label{energy}
{\cal E}_\a{}^i:=-\frac 14  e^i{}_\a( C_{ jk}{}^\b
{\cal H}^{jk}{}_\b) + C_{\a k}{}^\b {\cal H}^{ik}{}_\b\,.
\end{equation}
Here ${\cal H}^{ij}{}_\a$ are the gravitational {\it excitations} that
are in general linear in the field strengths $C^{ij}{}_\a$. Moreover,
the energy-momentum tensor density of matter is given by ${\cal
  T}_\a{}^i$, while ${\cal E}_\a{}^i$ has the interpretation of the
energy-momentum tensor density of the gravitational field. The
gravitational field equations (\ref{fieldeq}) resemble Maxwell's
equations. Indeed, the first set of equations expresses the divergence
of excitation in terms of the sources including gravity (${\cal
  T}_\a{}^i+{\cal E}_\a{}^i$); that is, all physical fields carry
energy-momentum. The second set of equations (\ref{fieldeq}) is
``solved'' by Eq.~(\ref{fieldstrength}), which expresses the field as
the (exterior) derivative of the potential.

To recover general relativity in this scheme, we must assume that
\begin{equation}\label{CcalH}
  {\cal H}^{ij}{}_\a=  \frac{\sqrt{-g}}{\kappa}\,\mathfrak{C}^{ij}{}_\a\,,
\end{equation}
where $\kappa=8\pi G$ and $G$ is Newton's constant of
gravitation. Here
\begin{equation}\label{frakC}
\mathfrak{C}_{ij}{}^\a =\frac 12\,
    C_{ij}{}^\a -C^\a{}_{[ij]}+2e_{[i}{}^\a C_{j]\g}{}^\g
\end{equation}
is the {\it modified} field strength.  It has been shown that Eqs.\
(\ref{fieldeq})-(\ref{frakC}) are an equivalent representation of
Einstein's theory of gravitation \cite{HehlHeld}.

4.\ {\it Nonlocal GR$_{||}$.} The translational gauge theory of
gravity bears a striking resemblance to electrodynamics. Just as
electromagnetism involves relations connecting excitation ${\cal
  H}^{ij} = (D, H)$ with the field strength $F_{ij} = (E, B)$, one may
look upon Eq.~(\ref{CcalH}) as a similar constitutive relation in
GR$_{||}$.  More generally, one can extend constitutive relations to
nonlocal ones; in particular, it is possible to imagine such relations
in vacuum in connection with acceleration-induced nonlocal
electrodynamics \cite{Muench00}. In a similar way, it is possible to
introduce nonlocality into the framework of teleparallel gravity.

Consider, for example, instead of Eq.~(\ref{CcalH}),
\begin{eqnarray}\label{excit3}
  {\cal H}^{ab}{}_c(x)&\!\!=\!\!&\frac{1}{\kappa}\sqrt{-g(x)}\,
  \lbrack\mathfrak{C}^{ab}{}_c(x)
    -\int \Omega^{ai}\Omega^{bj}\Omega_{ck} \,
  \nonumber\\ &&
\times {K}(x,y)
    \mathfrak{C}_{ij}{}^k(y)\sqrt{-g(y)}\,d^4 y\rbrack\,,
\end{eqnarray}
where $\Omega (x,y)$ is Synge's world-function \cite{Synge}, which is
half the square of the geodesic distance connecting $x$ and $y$. The
world-function and related bitensors are all smooth functions provided
we assume that in the spacetime region under consideration each pair
of events can be joined by a unique geodesic. In Eq.~(\ref{excit3}),
the coordinate indices $a, b, c,...$ refer to $x$, while $i,j,k,...$
refer to $y$; moreover, we define the bitensors
\begin{equation}\label{world1}
\Omega_a(x,y)=\frac{\partial\Omega}{\partial x^a}\,,\qquad
\Omega_i(x,y)=\frac{\partial\Omega}{\partial y^i}\,,
\end{equation}
and note that $\Omega$ satisfies the basic partial differential equations
\begin{equation}\label{world2}
2\Omega=g^{ab}\Omega_a\Omega_b=g^{ij}\Omega_i\Omega_j\,.
\end{equation}
The covariant derivatives at $x$ and $y$ commute for any bitensor and
it is possible to show that $\Omega_{ai} (x,y) = \Omega_{ia}(x,y)$ are
dimensionless bitensors such that
\begin{equation}\label{2point}
\lim_{y\rightarrow x}\Omega_{ai}(x,y)=-g_{ai}(x)\,.
\end{equation}

Equation (\ref{excit3}) introduces the causal scalar kernel $K(x,y)$
that indicates the nonlocal deviation from Einsteinian gravity in our
model. Here $K(x,y) = 0$ if $x$ lies in the past of $y$; in Minkowski
spacetime, for instance, this kernel would in general be nonzero only
for $x^0\ge y^0$. The kernel $K$ is in general a function of the
coordinate invariants that can be constructed at $x$ and $y$ within
Weitzenb\"ock spacetime; for instance, it could depend on the
Weitzenb\"ock invariants
\begin{equation}\label{Winvariants}
C_{ijk}C^{ijk}\,,\quad C_{kji}C^{ijk}\,,\quad C_{ij}{}^j C^{ik}{}_k\,.
\end{equation}
Other coordinate invariants can be formed, for example, from $\Omega$
and its covariant derivatives. As an illustration, let us consider
$\Omega^a e_a{}^\a(x)$ and $\Omega^i e_i{}^\a(y)$, which are
particularly interesting in view of the linear approximation scheme
that is developed below; indeed, in Minkowski spacetime these reduce
to $x^\a - y^\a$ and $y^\a - x^\a$, respectively.  We note that the
nonlocal constitutive relation (\ref{excit3}) is highly nonlinear in
the gravitational variables. Its substitution in the field equations
(\ref{fieldeq}) then results in a nonlocal generalization of
Einstein's theory of gravitation characterized by a given nonlocal
kernel $K(x, y)$. It should be emphasized that other nonlocal models
can be considered based on more complicated constitutive relations; in
fact, Eq.~(\ref{excit3}) represents the simplest nonlocal constitutive model
involving a scalar kernel. The physical origin of this kernel is
beyond the scope of our constitutive approach; a more basic theory is
needed for its determination.

Let us observe that our approach to nonlocal gravitation
differs essentially from other nonlocal modifications of general
relativity; see, for example, \cite{Soussa2003,Barvinsky2003} and the
references cited therein.

5.\ {\it Linear approximation.} The Weitzenb\"ock
spacetime in the $\Gamma_i{}^{\a\b}=0$ gauge reduces to Minkowski
spacetime for $e_i{}^\a=\d_i^\a$. Thus it is possible to develop a
linear approximation of our nonlocal theory of gravitation by assuming
\begin{equation}\label{lin2}
e_i{}^\a=\d_i^\a+\psi^\a{}_i\,,
\end{equation}
where the nonzero components of $\psi^\a{}_i$ have magnitudes that are
much smaller than unity. Henceforth $\psi^\a{}_i$ will be taken into
account at the first order of approximation, where the distinction
between holonomic coordinate indices and anholonomic tetrad indices
disappears. The metric of the Weitzenb\"ock spacetime is then given by
\begin{equation}\label{lin4}
g_{ij}=\eta_{ij}+h_{ij}\,,\qquad h_{ij}=2\psi_{(ij)}\,,
\end{equation}
and the gravitational field strengths can now be expressed as
\begin{eqnarray}
\label{lin3} C_{ij}{}^k&=&2\psi^k{}_{[j,i]}\,,\\
\nonumber\mathfrak{C}^{ij}{}_k&=&-\frac
12\left(h_{k\,\;,}^{\,\;i\;\,j}-h_{k\,\;,}^{\,\;j\;\,i}\right)+\psi^{[ij]}{}_{,k}\\
\label{lin5}&&+\, \d^i_k\left(\psi_,{}^j-\psi_{l\;\,,}^{\;\,j\;
    \,l}\right)-\d^j_k\left(\psi_,{}^{\,i}- \psi_{l\;\,,}^{\,\;i\;\,l}\right)\,,
\end{eqnarray}
where $\psi=\eta_{ij}\psi^{ij}$. Moreover, the constitutive relation
(\ref{excit3}) takes the form
\begin{eqnarray}\label{option1}
\kappa {\cal H}^{ij}{}_\alpha (x)= \mathfrak{C}^{ij}{}_\alpha (x)
                + \int {\cal K}(x, y) \mathfrak{C}^{ij}{}_\alpha (y) d^4y\,,
\end{eqnarray}
where ${\cal K}(x, y)$ is the scalar kernel $K(x, y)$ evaluated in the
Minkowski spacetime limit. Thus ${\cal K}$ is in general nonzero for
$x^0\ge y^0$. The substitution of these relations in
Eq.~(\ref{fieldeq}) results in
\begin{equation}\label{option2}
\partial_j\mathfrak{C}^{ij}{}_k+\int\frac{\partial
{\cal K}(x,y)}{\partial
x^j}\mathfrak{C}^{ij}{}_k(y)d^4y=\kappa T_k{}^i\,.
\end{equation}
This is the {\it main equation} of our linearized theory; for ${\cal
  K} = 0$, it reduces to linearized general relativity, since it is
simple to prove that the Einstein tensor in the general linear
approximation is given by
\begin{equation}\label{feq1}
  G^i{}_j=\partial_k\mathfrak{C}^{ik}{}_j\,.
\end{equation}
We note that $\partial_i T_k{}^i=0$ follows in general from the
nonlocal field equations (\ref{option2}). The constitutive relation
(\ref{option1}) is valid for sufficiently weak gravitational fields as
the constitutive kernel $K$ in the linear approximation becomes
completely independent of the gravitational field variables.

6.\ {\it Reciprocal kernel.} Let us now assume that ${\cal K}(x, y)$ is a
function of $x- y$; then, Eq.~(\ref{option2}) implies that
\begin{equation}\label{integraleq}
  G_{ij}(x)+\int{\cal K}(x-y)G_{ij}(y)d^4y=\kappa T_{ij}(x)\,,
\end{equation}
where we have neglected certain boundary terms, since the derivatives
of $\psi^k{}_i$ are expected to vanish at infinity. This is a Fredholm
equation of the second kind and it can be formally solved by the
Liouville-Neumann method of successive substitutions. That is, one can
move the second term from the left-hand side of Eq.~(\ref{integraleq})
to the right-hand side and then substitute for $G_{ij}(y)$ in the
integrand its value given by the resulting equation. The repetition of
this process would lead to an infinite series in terms of iterated
kernels ${\cal K}_n (x, y)$, $ n = 1, 2, 3, ...$ , that are in general
defined by $ {\cal K}_1(x,y)={\cal K}(x,y)$ and
\begin{equation}\label{ker1}
{\cal K}_{n+1}(x,y)=-\int {\cal K}(x,z){\cal K}_n(z,y)d^4z\,.
\end{equation}
If the resulting infinite series is uniformly convergent
\cite{Tricomi}, we can define a {\it reciprocal kernel} ${\cal R}(x,
y)$ given by
\begin{equation}\label{ker2}
-{\cal R}(x,y)=\sum_{n=1}^\infty {\cal K}_n(x,y)
\end{equation}
such that the formal solution of equation (\ref{integraleq}) can be written as
\begin{equation}\label{feq2}
  G_{ij}(x)=\kappa T_{ij}(x)+\kappa\int{\cal R}(x,y)T_{ij}(y)d^4y\,.
\end{equation}

To preserve causality, it is convenient to assume that the constitutive
kernel is  of the form
\begin{equation}\label{product}
{\cal K}(x-y)=\d (x^0-y^0)p(\mathbf{x}-\mathbf{y})\,.
\end{equation}
It follows from Eqs.~(\ref{ker1}) and (\ref{ker2}) that all of the
iterated kernels as well as the reciprocal kernel are of this same
form, since the corresponding spatial integration in Eq.~(\ref{ker1})
would be over the entire Euclidean space. In particular,
\begin{equation}\label{overline}
{\cal R}(x-y)=\d (x^0-y^0)q(\mathbf{x}-\mathbf{y})\,,
\end{equation}
where $p$ and $q$ are reciprocal spatial kernels.

7.\ {\it Dark matter.} It is possible to express Eq.~(\ref{feq2})
as
\begin{equation}\label{feqdark}
G_{ij}=\kappa\left(T_{ij}+\mathfrak{T}_{ij} \right)
\end{equation}
and interpret the new source for linearized Einsteinian gravity as
``dark matter''. Consequently, $\mathfrak{T}_{ij}$ is the symmetric
energy-momentum tensor of ``dark matter'' that is given by
\begin{equation}\label{dark2}
\mathfrak{T}_{ij}(x)=\int{\cal R}(x,y)T_{ij}(y) d^4y\,.
\end{equation}
Thus ``dark matter'' is in effect the integral transform of matter by
the reciprocal kernel ${\cal R}(x,y)$. In our simple model, the
characteristics of ``dark matter'' should be quite similar to actual
matter in accordance with Eq.~(\ref{dark2}). For example, the ``dark
matter'' associated with dust would be pressure-free, while
$\frak{T}_{ij}$ is traceless for radiation with $T_k{}^k =0$. In
particular, for a reciprocal kernel of the form (\ref{overline}), we
have for dust of density $\rho$,
\begin{equation}\label{density}
  \rho_{\rm D}(t,\mathbf{x})=\int
  q(\mathbf{x}-\mathbf{y})\rho(t,\mathbf{y})d^3y\,,
\end{equation}
so that the density of ``dark matter'' $ \rho_{\rm D}$ is in effect
the convolution of $\rho$ and $q$.

It appears natural to speculate that the astrophysical evidence for
dark matter may in fact be a manifestation of the nonlocal aspect of
classical gravity. In the linear approximation, the kernel in
Eq.~(\ref{dark2}) is universal in the sense that it is independent of
the particular configuration of matter under consideration. It follows
from the application of Eqs.~(\ref{feqdark}) and (\ref{dark2}) to the
solar system that our predicted dark matter must be a rather small
fraction of matter; otherwise, there would be conflict with the
solar-system tests of GR.  This requirement can be met with an
appropriate choice of kernel based on the observational evidence for
dark matter. Consider, for instance, the circular motion of stars in
the disk of a spiral galaxy in connection with the problem of dark
matter in such galaxies
\cite{Rubin,Roberts,Sofue,Carignan,Jacob2004,Jacob2006,Bruneton2008,TohlineKuhn,Jacob1988}. Outside
the bulge, the Newtonian acceleration of gravity for each star at
radius $|\mathbf{x}|$ is toward the galactic center with magnitude
$v_0^{\,2}/|\mathbf{x}|$, where $v_0$ is the (approximately) constant
speed of stars in conformity with the observed rotation curves of
spiral galaxies \cite{Rubin,Roberts,Sofue,Carignan}. The density of
dark matter in the disk responsible for this behavior is given by
$v_0^{\,2}/\left(4\pi G|\mathbf{x}|^2\right)$, which follows from
Poisson's equation. Neglecting the dimensions of the galactic bulge
and setting $\rho(t,\bfy)=M\d(\bfy)$, where $M$ is the effective mass
of the galaxy, Eq.~(\ref{density}) then implies that
\begin{equation}\label{q(x)}
q(\bfx)=\frac{1}{4\pi\lambda}\frac{1}{|\bfx|^2}\,,
\end{equation}
where $\lambda=GM/v_0^{\,2}$ is a constant length parameter of order 1
kpc. With this reciprocal kernel, the Newtonian limit of our
linearized nonlocal theory is given by
\begin{equation}\label{79}
  \nabla^2  \Phi=4\pi G\left[\rho(t,\bfx)+\frac{1}{4\pi\lambda}
    \int\frac{\rho(t,\bfy)d^3y}{|\bfx-\bfy|^2}\right]\,,
\end{equation}
where $\Phi$ is the Newtonian potential. For a point mass $m$ with
$\rho(t,\bfx)=m\d(\bfx)$, Eq.~(\ref{79}) has the solution
\begin{equation}\label{80}
  \Phi(t,\bfx)=-\frac{Gm}{|\bfx|}+\frac{Gm}{\lambda}
  \ln\left(\frac{|\bfx|}{\lambda}\right)\,,
\end{equation}
where the logarithmic term due to dark matter has a negligible
influence in the solar system, as required.

Remarkably, Eqs.~(\ref{79}) and (\ref{80}) coincide with the
Tohline-Kuhn scheme \cite{TohlineKuhn} that, apart from a disagreement
with the empirical Tully-Fisher law, has been quite successful in
dealing with dark-matter issues in galaxies and clusters---see the
lucid review by Bekenstein \cite{Jacob1988}. The universality of
kernel (\ref{q(x)}) implies $M\propto v_0^{\,2}$; therefore, the main
shortcoming of the linear approximation that results in a universal
kernel is that the Tully-Fisher relation ($M\propto v_0^{\,4}$) is
violated \cite{Jacob1988}.  \vspace{-1pt}

In a general treatment, an equation of the form (\ref{dark2}) may
still be pertinent, but with a kernel that would strongly depend on
the particular matter distribution. Furthermore, it would be
interesting to have a theory that would be able to predict the kernel
from first principles. But the development of such a theory remains a
task for the future as it is beyond the scope of the constitutive
approach adopted in this work.

{\it Acknowledgments.}  We wish to express our deep appreciation to
Jacob Bekenstein (Jerusalem) for constructive comments that helped
improve our paper. We are also grateful to Yakov Itin (Jerusalem),
Jos\'e Maluf (Brasilia), Eckehard Mielke (Mexico City), and Yuri
Obukhov (Moscow) for most useful remarks.

\end{document}